\documentclass[12pt]{article}
\usepackage{float}
\usepackage[dvips]{graphicx}
\usepackage{amsbsy}
\usepackage{amssymb}
\usepackage{color}
\usepackage{epic}

\newcommand{\beq}{\begin{equation}}
\newcommand{\eeq}{\end{equation}}
\newcommand{\bdm}{\begin{displaymath}}
\newcommand{\edm}{\end{displaymath}}
\newcommand{\beqr}{\begin{eqnarray}}
\newcommand{\eeqr}{\end{eqnarray}}
\newcommand{\beqrn}{\begin{eqnarray*}}
\newcommand{\eeqrn}{\end{eqnarray*}}
\def\l{\lambda}

\def\bchi{\boldsymbol{\chi}}

\def\ve{\varepsilon}
\def\a{\alpha}

\def\l{\lambda}

\begin{document}

\title{On the generating function of weight multiplicities for the representations of the Lie algebra $C_2$}

\author{Jos\'e Fern\'andez-N\'u\~{n}ez$^{\dagger}$, Wifredo Garc\'{\i}a-Fuertes$^{
\ddagger}$\\
\small Departamento de F\'\i sica, 
Facultad de Ciencias, Universidad de Oviedo, 33007-Oviedo, Spain\\
\small {\it $^\dagger$nonius@uniovi.es; $^\ddagger$wifredo@uniovi.es}\\
\and
 Askold M. Perelomov\\
\small Institute for Theoretical and Experimental Physics, Moscow, Russia.\\
\small {\it  aperelomof.uo@uniovi.es}}

\date{ }

\maketitle

\begin{abstract}\noindent
We use the generating function of the characters of $C_2$ to obtain a generating function for the multiplicities of the weights entering in the irreducible representations of that simple Lie algebra. From this generating function we derive some recurrence relations among the multiplicities and a simple graphical recipe to compute them. 
\end{abstract}
\bigskip


\medskip

\vfill\eject 
\section{Introduction}

Each irreducible representation of a simple Lie algebra is defined by a set of weights which, for rank two algebras, can be conveniently arranged in a two-dimensional weight diagram. These weights result from successive applications of the lowering operators $E_{-\alpha}$ corresponding to the positive roots of the algebra to the highest weight of the representation. As there are, in general, several ways by which a particular weight can be obtained in this form, the weights forming the representation enter in it with some multiplicity. The computation and understanding of weight multiplicities has been a subject of much research along the years \cite{wi37}--\cite{mopa82} and, as it is a rule when dealing with Lie algebra representations, one of the most efficient tools available to address the question is the theory of characters. In a recent paper \cite{nfp14}, we have presented a general method for computing the generating function of the characters of simple Lie algebras which is based on the theory of the quantum trigonometric Calogero-Sutherland system \cite{ca71}--\cite{op76} (see also \cite{ps79,ow07} for other approaches to that problem). In particular, we have applied the method to the cases of the Lie algebras $A_2$ and $C_2$. The aim of this note is to supplement the results of \cite{nfp14} by showing how they can be used to obtain some useful generating functions for weight multiplicities. In doing so, we will specialize to the case of the algebra $C_2$, given that the case of the generating function for multiplicies of $A_2$ has been soundly treated in reference \cite{bgw68}. 

Let us recall, to begin with, the way in which characters and weight multiplicities are related. Let ${\cal A}$ be a simple Lie algebra of rank $r$ with fundamental weights 
$\lambda_1,\lambda_2,\ldots,\lambda_r$ and let us denote $R_{\lambda}$ the irreducible 
representation of ${\cal A}$ with highest weight 
$\lambda=p_1 \lambda_1+p_2 \lambda_2+\cdots+p_r \lambda_r$. 
The character of this representation is defined as
\bdm
\bchi_{p_1,p_2,\dots,p_r}=\sum_{w} \mu_{w} e(w)
\edm
where  the sum extends to all weights $w$ entering in the representation, $\mu_{w}$ is the 
mutiplicity of the weight $w$ and, if $w=m_1 \lambda_1+m_2 \lambda_2+\cdots+m_r \lambda_r$, 
then $e(w)$ is 
\bdm
e(w)=\exp\Big(i\sum_{l=1}^r m_l \varphi_l\Big)=x_1^{m_1} x_2^{m_2}\cdots x_r^{m_r}, 
\edm
where $\varphi_1,\varphi_2,\ldots,\varphi_r$ are angular coordinates on the maximal torus and  
$x_l$ are complex phases, \mbox{$x_l=e^{i \varphi_l}$}. The multiplicity $\mu_{p_1,p_2,\dots,p_r}(m_1,m_2,\dots,m_r)$ of the weight $w\equiv(m_1,m_2,\dots,m_r)$ in the representation $R_\lambda$, $\l\equiv(p_1,p_2,\dots,p_r)$, can be computed as
\beqr
&&\mu_{p_1,p_2,\dots,p_r}(m_1,m_2,\dots,m_r)\nonumber\\
&&\qquad\qquad=\frac{1}{(2\pi)^r}\int_0^{2\pi}
d\varphi_1 e^{-i m_1 \varphi_1}\int_0^{2\pi}d\varphi_2e^{-i m_2 \varphi_2}\ldots\int_0^{2\pi}
d\varphi_r e^{-i m_r \varphi_r}\bchi_{p_1,\dots,p_r}\nonumber\\
[2pt]
&&\qquad\qquad\,=\frac1{(2\pi i)^r}\oint d x_1\oint d x_2\ldots\oint d x_r \frac{\bchi_{p_1,\dots,p_r}}
{x_1^{1+m_1} x_2^{1+m_2}\cdots x_r^{1+m_r}}\, , \label{eq:nmp}
\eeqr
where the integrals in the second line are along the unit circles on the $r$ complex planes 
parametrized by the complex coordinates $x_1,x_2,\ldots,x_r$. 

In view of (\ref{eq:nmp}), the 
generating function for the multiplicities of the weight $w$ in all the representations of ${\cal A}$
\beq
\label{amne}
A_{m_1,m_2,\dots,m_r}(t_1,t_2,\ldots,t_r)=\sum_{p_1=0}^\infty\sum_{p_2=0}^\infty\cdots
\sum_{p_r=0}^\infty t_1^{p_1} t_2^{p_2}\cdots t_r^{p_r}\mu_{p_1,\dots, p_r}(m_1,\dots,m_r)
\eeq
comes from the formula
\beq
\label{amni}
A_{m_1,\dots,m_r}(t_1,t_2,\ldots,t_r)=\frac{1}{(2\pi i)^r}\oint d x_1
\oint d x_2\ldots\oint d x_r \frac{G(t_1,t_2,\ldots,t_r ; z_1,z_2,\ldots,z_r)}
{x_1^{1+m_1} x_2^{1+m_2}\cdots x_r^{1+m_r}} ,\label{eq:igm}
\eeq
where $G(t_1,t_2,\ldots,t_r ; z_1,z_2,\ldots,z_r)$ is the generating function of the characters
\bdm
G(t_1,t_2,\ldots,t_r ; z_1,z_2,\ldots,z_r)=\sum_{p_1=0}^\infty\sum_{p_2=0}^\infty\cdots\sum_{p_r=0}^\infty t_1^{p_1} t_2^{p_2}
\cdots t_r^{p_r} \bchi_{p_1,\dots,p_r}(z_1,z_2,\ldots,z_r)  \label{eq:gcar}
\edm
and we have chosen to express the latter by means of a set of variables $z_1,z_2,\ldots,z_r$ which coincide with the characters of the representations corresponding to the fundamental weights. Even for low-rank algebras and small values of the indices $m_j$ the integrand in (\ref{eq:igm}) is a quite complicated rational function but, nevertheless, the integral can be evaluated by iterated application of the Cauchy's residue theorem in each complex plane. 

Let us consider, for instance, the case of the generating function of zero weight multiplicities for the Lie algebra $A_2$. According to \cite{ov90}, see also \cite{nfp14}, the fundamental characters are 
\bdm
z_1=x_1+\frac1{x_2}+\frac{x_2}{x_1}\,, \quad z_2=x_2+\frac1{x_1}+\frac{x_1}{x_2}\,,
\edm
whereas the generating function $G$ is \cite{nfp14}
\beqrn
G(t_1,t_2;z_1,z_2)&=&\frac{1-t_1 t_2}{(1-t_1 z_1+t_1^2 z_2-t_1^3)(1-t_2 z_2+t_2^2 z_1-t_2^3)} \\
[2pt]
&=&\frac{(1 - t_1 t_2)\, x_1^2\, x_2^2}{
    (t_2 - x_1) ( t_1 x_1-1) (t_1 - x_2) (t_2 x_1 - x_2) (t_1 x_2-x_1) (t_2 x_2-1)}.
\eeqrn
Then we have to compute
\bdm
A_{0,0}(t_1,t_2)=\frac{1}{(2\pi i)^2}\oint d x_1 \oint d x_2 
\frac{(1 - t_1 t_2) x_1 x_2}{
    (t_2 - x_1) ( t_1 x_1-1) (t_1 - x_2) (t_2 x_1 - x_2) (t_1 x_2-x_1) (t_2 x_2-1)}
\edm
and we choose to perform the $x_1$ integral first. As $|x_1|=|x_2|=1$ and $t_1,t_2<1$, there are poles inside 
the unit circle for $x_1=t_2$ and $x_1=t_1 x_2$. Thus, by computing the residues, we find
\beqrn
J_1(t_1,t_2;x_2)=\frac{1}{2\pi i}\oint d x_1 \frac{G(t_1,t_2;z_1,z_2)}{x_1 x_2}
=\frac{(1 + t_1 t_2) x_2}{(t_1 - x_2) 
(t_2^2 - x_2) (t_1^2 x_2-1) (t_2 x_2-1)}.
\eeqrn
Now, integrating $J_1(t_1,t_2;x_2)$, which has poles inside the $x_2$ unit circle at $x_2=t_1$ and $x_2=t_2^2$,   
we finally obtain the generating function for zero weight multiplicities as
\beqrn
A_{0,0}(t_1,t_2)=\frac{1}{2\pi i} \oint d x_2J_1(t_1,t_2,x_2)=\frac{1 - t_1^3\, t_2^3}{(1 - t_1^3) (1 - t_1 t_2)^2 (1 - t_2^3)}\,.
\eeqrn

\section{The generating function $A_{m,n}(t_1,t_2)$ for $C_2$}
In the case of $C_2$, the fundamental characters are \cite{ov90}
\beq
\label{eq:b2}
z_1=x_1 +\frac{1}{x_1}+\frac{x_1}{x_2}+\frac{x_2}{x_1}\,,\quad z_2
=1+x_2+\frac{1}{x_2}+\frac{x_1^2}{x_2}+\frac{x_2}{x_1^2} \,,
\eeq
and the generating function of the characters is \cite{nfp14}
\beqr
&&G(t_1,t_2;z_1,z_2)\nonumber\\
&&=\frac{1+t_2-z_1 t_1 t_2+t_1^2 t_2+t_1^2 t_2^2}{(1-(t_1+t_1^3) z_1+t_1^2 (z_2+1)+t_1^4)(1-(t_2+t_2^3)(z_2-1)+t_2^2(z_1^2-2z_2)+t_2^4)}\nonumber\\ 
[2pt]
&&=\frac{x_1^3 x_2^2 ((1 + t_2) x_1 x_2 + t_1^2 t_2 (1 + t_2) x_1 x_2 - 
      t_1 t_2 (1 + x_2) (x_1^2 + x_2))}
  {(x_1-t_1) (t_1 x_1-1) (t_1 x_1 - x_2) (t_2 x_1^2 - x_2) 
    (x_2-t_2) (x_1 - t_1 x_2) (x_1^2 - t_2 x_2) (t_2 x_2-1)}\ . \label{eq:gcb2}
\eeqr
Thus, for $m_1=m_2=0$, the poles of integrand $G/x_1x_2$  in (\ref{eq:igm}) are easy to identify, and going through the steps seen in the previous example, we eventually find that the generating function for zero-weight multiplities for $C_2$ is
\bdm
A_{0,0}(t_1,t_2)=\frac{1 + t_1^2 t_2}
{(1 - t_1^2)^2 (1 - t_2) (1 - t_2^2)}\,.
\edm

The form of $A_{0,0}(t_1,t_2)$ is simple enough to allow us to go one step further. We can expand $A_{0,0}(t_1,t_2)$ as a sum of partial fractions
\[
A_{0,0}(t_1,t_2)=\frac12\left[\frac1{(1-t_1^2)(1-t_2^2)}+\frac{1+t_1^2}{(1-t_1^2)^2(1-t_2)^2}\right]
\]
whose Taylor series are quite simple. Matching coefficients yields the general formula for the multiplicities $\mu_{p,q}(0,0)$ of the zero weight as
\[
\mu_{p,q}(0,0)=\frac12\ve_p[\ve_q+(p+1)(q+1)]
\]
where $\ve_p=1$ for $p$ even, or $\ve=0$ for $p$ odd.

The calculations needed to obtain the generating functions for the multiplicities of other low-lying weights go along the same lines and we list some results in the Appendix. However, using directly formula (\ref{amni}) to find the generating function of the multiplicities of a general weight $(m,n)$ seems to be quite involved. In order to make progress, it it is more convenient to introduce  a new generating function $H(t_1,t_2;y_1,y_2)$ defined as
\bdm
H(t_1,t_2;y_1,y_2)=
\sum_{m=0}^\infty\sum_{n=0}^\infty\sum_{p=0}^\infty\sum_{q=0}^\infty \mu_{p,q}(m,n) y_1^m y_2^n t_1^p t_2^q 
\edm
which collects the multiplicities $\mu_{p,q}(m,n)$ of all weights $m\lambda_1+n\lambda_2$ in all the 
representations $R_{p\lambda_1+q\lambda_2}$ of $C_2$. Expressing $\mu_{p,q}(m,n)$ as was done in (\ref{eq:nmp}), the sums in the indices $m$ and $n$ yield geometric series, leading to the formula 
\bdm
H(t_1,t_2;y_1,y_2)=\frac{1}{(2\pi i)^2}\oint d x_1\oint d x_2 
\frac{G(t_1,t_2;z_1,z_2)}{(x_1-y_1)(x_2-y_2)}\ ,
\edm
which, after substitution of  (\ref{eq:b2}) and (\ref{eq:gcb2}), takes the form of a rational integral to be evaluated by means of Cauchy's theorem as in the previous examples. The result is
\beq
H(t_1,t_2;y_1,y_2)=\frac{a+b_1 y_1+b_2 y_2+c_{1,2}y_1 y_2+d y_1^2 +e y_1^2 y_2}
{(1 - t_1^2)^2(1 - t_2^2)(1 - t_2)(1 - t_1y_1)(1 - t_2^2y_1^2)(1 - t_1^2y_2)(1 - t_2y_2)}
\label{eq:gengen}
\eeq
where
\beqrn
a&=&1+t_1^2 t_2,\hspace{2cm}b_1=t_1 t_2(1-t_1^2),\hspace{1.3cm} b_2=-t_1 t_2(t_1^3+t_1 t_2),\\
c_{1,2}&=&t_1 t_2 (t_1^4-t_1^2),\hspace{1cm} d=-t_1^2 t_2^2(1+t_2),\hspace{1cm}e
=t_1^2 t_2^2(t_1^2+t_1^2 t_2+t_2^2-1).
\eeqrn
Now, trading the factors $(1 - t_1y_1)(1 - t_2^2y_1^2)(1 - t_1^2y_2)(1 - t_2y_2)$ in the denominator by geometric series in $y_1$ and $y_2$, we can rewrite $H(t_1,t_2;y_1,y_2)$ as a series
\bdm
H(t_1,t_2;y_1,y_2)=\sum_{m=0}^\infty\sum_{n=0}^\infty A_{m,n}(t_1,t_2) y_1^m y_2^n
\edm
such that the coefficients 
are precisely the generating functions for weight multiplicities which we are seeking for. From (\ref{eq:gengen}), and after some tedious 
algebra, one can obtain  the explicit form of these generating functions as
\beqr
\label{amn}
&&A_{m,n}(t_1,t_2)\nonumber\\
[2pt]
&&=\frac{t_1^{m+2n+2} (t_1^2-t_2^2)(1-t_2^2)-t_1^{m+2} t_2^{n+1} 
(1-t_1^2)(1-t_2^2)-t_2^{m+n+1}(t_1^2-t_2)(1-t_1^2) f(t_1,t_2)}{(1-t_1^2)^2(1-t_2^2)(1-t_2)(t_1^2-t_2^2)(t_1^2-t_2)}
\eeqr
with
\bdm
f(t_1,t_2)=\left\{\begin{array}{ll}t_1^2+t_2\,,& {\rm for}\ \ m\ \  
{\rm even}\\t_1(1+t_2)\,,& {\rm for}\ \ m\ \  {\rm odd}\end{array}\right. 
\edm
thus generalizing the results for low-lying multiplicities explained before.\footnote{After this work was completed we learned about the very interesting paper by Dokovi\'c \cite{dok} in which he obtains the generating functions for weight multiplicities for the simple Lie algebras of rank 2. Unfortunately, the result quoted in that paper for $B_2\equiv C_2$ and $m$ even is not correct.}

This expression of the generating function differs from the examples of the Appendix by the factors $(t_1^2-t_2^2)(t_1^2-t_2)$ in the denominator. In fact, some further simplification work shows that these factors cancel out, giving
\beqrn
A_{m,n}(t_1,t_2)&=&\frac{1}{D}\Big[(1-t_2^2)\sum_{j=0}^{n-1} t_1^{m+2n-2j} t_2^j+(1-t_1^2+t_2-t_2^3)\sum_{j=0}^{m\over 2} t_1^{2j} t_2^{m+n-2j}\\
&-&(1-t_1^2-t_2^2) t_2^{m+n+1}+t_1^{m+2} t_2^n\Big]
\eeqrn
for even $m$ and
\beqrn
A_{m,n}(t_1,t_2)&=&\frac{1}{D}\Big[(1-t_2^2)\sum_{j=0}^{n-1} t_1^{m+2n-2j} t_2^j
+\,(1+t_2-t_2^2-t_1^2t_2)\sum_{j=0}^{{m-3}\over 2} t_1^{2j+1} t_2^{m+n-2j-1}\\
&+&(1+t_2-t_2^2) t_1^m t_2^n+t_1 t_2^{m+n+1}\Big]
\eeqrn
for odd $m$, with $D=(1-t_1^2)^2(1-t_2^2)(1-t_2)$.

After the generating functions are known some other interesting results come from them. In particular, looking at their form for low $m$ and $n$, one can identify two different recurrence relations among the multiplicities $\mu_{p,q}(m,n)$, and with some additional labour, it is possible to show that these recurrence relations are valid in general. This is described in the next two sections.
 \section{The first recurrence relation}
The first recurrence relation is among the multiplicities 
of a fixed weight $m\l_1+n\l_2$ in different representations. Let us call 
\beq
X_{m,n}(t_1,t_2)=(1-t_1^2)(1-t_2) A_{m,n}(t_1,t_2).
\eeq
From the definition of $A_{m,n}(t_1,t_2)$, one has
\bdm
X_{m,n}(t_1,t_2)=\sum_{p=0}^\infty\sum_{q=0}^\infty [\mu_{p,q} (m,n)-\mu_{p-2,q} (m,n)-\mu_{p,q-1} (m,n)+\mu_{p-2,q-1} (m,n)]t_1^p\, t_2^q 
\edm
while the explicit expressions given above yield
\beqrn
X_{m,n}(t_1,t_2)&=&\frac{(1-t_2^2){\displaystyle\sum_{j=0}^{n-1}}\,t_1^{m+2n-2j} t_2^j+(1-t_2^2)\Big[ (1+t_2)\displaystyle\sum_{j=0}^{\frac{m}{2}}\,t_1^{2j}t_2^{m+n-2j}-t_2^{m+n+1}\Big]}{(1-t_1^2)(1-t_2^2)}\\
&+&\frac{t_1^2 t_2^{m+n+1}+t_2^{m+n+2}}{(1-t_1^2)(1-t_2^2)}
\eeqrn
for $m$ even and
\beqrn
X_{m,n}(t_1,t_2)&=&\frac{(1-t_2^2)\displaystyle\sum_{j=0}^{n-1} t_1^{m+2n-2j} t_2^j+(1-t_2^2)(1+t_2)\Big[\sum_{j=0}^{\frac{m-3}{2}}t_1^{2j+1}t_2^{m+n-1-2j}+t_1^mt_2^n\Big]}{(1-t_1^2)(1-t_2^2)}\\
&+&\frac{(1+t_2)t_1 t_2^{m+n+1}}{(1-t_1^2)(1-t_2^2)}
\eeqrn
for $m$ odd. 

Let us consider the formula for $m$ even and compare it with the diagram 
of Figure 1, which shows all representations $R_{p\l_1+q\l_2}$ with nonzero 
multiplicity for a weight $m\l_1+n\l_2$. The labels   $(m,n)$, $(p,q)$ etc., represent coordinates in the non-Euclidean $(\l_1$,$\l_2$)-plane. The first term in $X_{m,n}(t_1,t_2)$ 
can be expanded as a geometric series which contains, always with coefficient equal to one, 
all the products $t_1^p t_2^q$ for $p$ and $q$ corresponding to points in the diagonals 
beginning in the segment $AB$. In a similar way, the second term in $X_{m,n}(t_1,t_2)$ 
gives all such products for the points in the diagonals normal to the line from $A$ to $C$, 
the third corresponds to the diagonals beginning in $D^\prime, E^\prime, F^\prime,\ldots$, 
etc, and the fourth to the diagonals from $D,E,F,\ldots$ etc. From this and an analogous 
analysis for $m$ odd, we can finally conclude that
\beq
\label{recrel}
\mu_{p,q} (m,n)-\mu_{p-2,q} (m,n)-\mu_{p,q-1} (m,n)+\mu_{p-2,q-1} (m,n)=y_{p,q}(m,n) \label{eq:rec1}
\eeq
where $y_{p,q}(m,n)=1$ if $(p,q)$ labels a irreducible representation of $C_2$ containing 
the weight $m\l_1+n\l_2$ (except for $m$ even, $p=0$ and $q$ of opposite parity to $n$, which gives    $y_{p,q}(m,n)=0$) and  $y_{p,q}(m,n)=0$ if the weight $m\l_1+n\l_2$ 
is not in $R_{p\l_1+q\l_2}$.
\medskip

\noindent{\bf Remark.} A geometric interpretation  can be given to the function $X_{m,n}(t_1,t_2)$ taking into account that, after direct substitution of the expression (\ref{amne}) of the generating function $A_{m,n}$, we can write it as a combination of sums of infinite geometric series, namely
\beqrn
X_{m,n}(t_1,t_2)&=&\sum_{j,k=0}^\infty t_1^{(m+2n)+2j-2k}t_2^k-\sum_{j,k=0}^\infty t_1^{(m-2)-2j-2k}t_2^{(n+1)+2j+k}\\
& -&\sum_{j,k=0}^\infty t_1^{-2j}t_2^{(m+n+1)+2j+2k}-\sum_{j,k=0}^\infty t_1^{-2-2j}t_2^{(m+n+2)+2j+2k}.
\eeqrn

Each term represents the contribution of the points, with coefficient +1 or $-1$, in a bidimensional lattice obtained after translating a point $(p_0,q_0)$ along some independent directions in the ($\l_1$,$\l_2$)-plane. For instance, the first sum represents the total contribution to $X_{m,n}$ of the lattice generated  from the basis $\{2\l_1,-2\l_1+\l_2\}$, with origin  the point $(m+2n,0)$ and nonnegative integers $j,k$. All the points thus generated have coefficient +1. With reference to the example in Figure 1, this is the lattice obtained by translating first the point $B(m+2n,0)$ along the $\l_1$-axis with step $2\l_1$; the linear lattice is then translated along $BA$ with step $-2\l_1+\l_2$.

The remaining contributions have the same interpretation, differing only in the fact that the coefficient is now $-1$; this means that points obtained from two contributions of different sign are superposed to give a null contribution. This way of counting gives the same lattice (Figure 1) as before,  for both $m$ even or odd.

\begin{figure}
\begin{center}
\setlength{\unitlength}{.7mm}
\begin{picture}(160,164)(0,0)

\put(0,0){\thicklines\line(1,0){160}}
\put(0,0){\thicklines\line(0,1){170}}
\put(0,0){\line(1,1){160}}
\put(0,0){\thicklines\vector(0,1){10}}
\put(0,0){\thicklines\vector(1,1){5}}
\put(0,100){\line(1,-1){30}}
\put(30,70){\line(1,0){40}}
\dashline{2}(0,40)(30,70)
\dashline{2}(30,61)(30,30)

\multiput(0,100)(0,10){7}{\circle*{2}}
\multiput(10,90)(0,10){8}{\circle*{2}}
\multiput(20,80)(0,10){9}{\circle*{2}}
\multiput(30,70)(0,10){10}{\circle*{2}}
\multiput(40,70)(0,10){10}{\circle*{2}}
\multiput(50,70)(0,10){10}{\circle*{2}}
\multiput(60,70)(0,10){10}{\circle*{2}}
\multiput(70,70)(0,10){10}{\circle*{2}}
\multiput(80,80)(0,10){9}{\circle*{2}}
\multiput(90,90)(0,10){8}{\circle*{2}}
\multiput(100,100)(0,10){7}{\circle*{2}}
\multiput(110,110)(0,10){6}{\circle*{2}}
\multiput(120,120)(0,10){5}{\circle*{2}}
\multiput(130,130)(0,10){4}{\circle*{2}}
\multiput(140,140)(0,10){3}{\circle*{2}}
\multiput(150,150)(0,10){2}{\circle*{2}}
\multiput(160,160)(0,10){1}{\circle*{2}}

\put(1,10){$\lambda_2$}
\put(5,1.5){$\lambda_1$}
\put(-30,99){\small$C(0,n+m)$}
\put(-6,120){\small$D$}
\put(-6,140){\small$E$}
\put(-6,160){\small$F$}
\put(12,120){\small$D'$}
\put(12,140){\small$E'$}
\put(12,160){\small$F'$}
\put(27,64){\small$A(m,n)$}
\put(69,64){\small$B(m+2n,0)$}
\put(29,26){\small$m$}
\put(-4,39){\small$n$}

\put(60,10){\thinlines\vector(1,0){10}} \put(71,10){$\a_1$}
\put(60,10){\thinlines\vector(-1,1){10}} \put(71,21){$\a_4$}
\put(60,10){\thinlines\vector(0,1){10}} \put(60,21){$\a_3$}
\put(60,10){\thinlines\vector(1,1){10}} \put(48,21){$\a_2$}
\end{picture}
\end{center}
\medskip

{\footnotesize Figure 1. The highest weights (modulo  Weyl reflections) corresponding 
to representations of $C_2$ containing the weight $m\l_1+n\l_2$ ($m=6$, $n=4$ in the example). Here, $\a_1,\dots,\a_4$ are the positive roots and $\l_1,\l_2$ are the fundamental weights, of magnitude $|\l_1|=1$ and $|\l_2|={\sqrt{2}}$).}

\end{figure}

\section{The second recurrence relation}
There is a second recurrence relation, this time among the multiplicities of weights of the 
same representation. Let  $P_{m,n}(t_1,t_2)$ be the function defined as
\beq
P_{m,n}(t_1,t_2)=A_{m,n}(t_1,t_2)-A_{m+2,n}(t_1,t_2)-A_{m,n+1}(t_1,t_2)+A_{m+2,n+1}(t_1,t_2).
\eeq
Then, from the previous expressions of $A_{m,n}(t_1,t_2)$, one can obtain
\bdm
P_{m,n}(t_1,t_2)=\frac{1}{1-t_2}\sum_{j=0}^n t_1^{m+2n-2j} t_2^j+\frac{1}{1-t_2}\sum_{j=0}^{\frac{m}{2}-1} t_1^{2j} t_2^{m+n-2j}-\frac{t_2^{m+n+1}}{(1-t_1^2)(1-t_2)}
\edm
for $m$ even and
\bdm
P_{m,n}(t_1,t_2)=\frac{1}{1-t_2}\sum_{j=0}^n t_1^{m+2n-2j} t_2^j+\frac{1}{1-t_2}\sum_{j=0}^{\frac{m-3}{2}} t_1^{2j+1} t_2^{m+n-1-2j}-\frac{t_1 t_2^{m+n+1}}{(1-t_1^2)(1-t_2)}
\edm
for $m$ odd. Expanding the denominators as geometric series and using de definition  of $A_{m,n}(t_1,t_2)$, 
one finds that for both parities
\beq
\mu_{p,q}(m,n)-\mu_{p,q}(m+2,n)-\mu_{p,q}(m,n+1)+\mu_{p,q}(m+2,n+1)= \varepsilon_{p,q}(m,n) \label{eq:rec2}
\eeq
where the right-hand member is a sum of three terms, $\varepsilon_{p,q}(m,n)=X+Y-Z$, 
which are zero except for the cases
\beqrn
X=1\ \  &{\rm when}& m\leq p\ \ {\rm and }\ \ \ p\leq m+2n \leq p+2 q\\
Y=1\ \  &{\rm when}& m\geq p+2\ \ {\rm and }\ \ \ m+n \leq p+q\\
Z=1\ \  &{\rm when}& m+n \leq q-1 .
\eeqrn

As $X=1$ and $Y=1$ do not occur at the same time, $\varepsilon_{p,q}(m,n)$ is always 1, 0 or $-1$. To describe the domains $D1$, $D0$ and $D(-1)$ in which $\varepsilon_{p,q}(m,n)$ takes these values, let us consider the weight diagram of the representation 
$p\l_1+q\l_2$ using now Cartesian coordinates, denoted as $[x,y]$ with $x=m$, $y=m+2n$, 
instead of the $m$ and $n$ labels of the weights. The weights entering in the diagram form a square lattice with a spacing equal to $|\l_2|$ which includes the highest weight of the 
representation and is contained in the polygon of vertices $[0, p+2q]$, $[p,p+2q]$, $[p+q,p+q]$ 
and $[0,0]$. If we call $P_X$, $P_Y$ and $P_Z$ the regions of the diagram in which, 
respectively, $X=1$, $Y=1$ and $Z=1$, it follows that:
\begin{itemize}
\item If $2q-2<p$, $P_X \cup  P_Y$ does not intersect $P_Z$. Thus, $D1=P_X\cup P_Y$, 
$D(-1)=P_Z$, and $D0$ the remaining weights. So, in this case $D1$ is the upper region 
of the diagram, starting from the horizontal line $y=p$, $D0$ is the area below that line 
and above the diagonal $y=2q-2-x$, and $D(-1)$ includes that diagonal and the weights below it, 
see Figure 2.
\item If $p\leq 2q-2\leq 2p$ the intersection $P_X\cap P_Z$ is the (possibly degenerate) 
triangle $T$ of vertices $[0,2q-2]$, $[0,p]$ and $[2q-2-p,p]$. In this case $D1=P_X\cup P_Y-T$, 
$D(-1)=P_Z- T$, and $D0$ the remaining weights. Now, as one can see in Figure 3, some weights 
above $y=p$ located  near the $y$ axis are in D0 instead of in D1.
\item If $2q-2 > 2p$ the intersection $(P_X\cup P_Y)\cap P_Z$ is the cuadrilateral $K$ 
of vertices $[0,p]$, $[0,2q-2]$, $[q-1,q-1]$ and $[p,p]$. Therefore $D1=P_X\cup P_Y-C$, 
$D(-1)=P_Z-C$, and $D0$ the remaining weights, see Figure 4.

\begin{figure}
\begin{center}
\setlength{\unitlength}{.7mm}
\begin{picture}(80,100)(0,0)
\put(0,0){\thicklines\line(1,0){80}}
\put(0,0){\thicklines\line(0,1){120}}
\put(0,0){\line(1,1){77}}
\put(0,0){\thicklines\vector(0,1){10}}
\put(0,0){\thicklines\vector(1,1){5}}
\put(50,100){\line(1,-1){27}}
\put(0,100){\line(1,0){50}}

\put(0,0){\circle{3}} \put(0,0){\circle*{2}}
\multiput(0,10)(10,0){2}{\circle{3}} \multiput(0,10)(10,0){2}{\circle*{2}}
\multiput(0,20)(10,0){3}{\circle{3}} \multiput(0,20)(10,0){3}{\circle*{2}} 
\multiput(0,30)(10,0){2}{\circle{3}} \multiput(0,30)(10,0){2}{\circle*{2}} 
\multiput(20,30)(10,0){2}{\circle{2}}
\put(0,40){\circle*{2}}\put(0,40){\circle{3}}
\multiput(10,40)(10,0){4}{\circle{2}} \multiput(0,50)(10,0){6}{\circle*{2}}
\multiput(0,60)(10,0){7}{\circle*{2}} \multiput(0,70)(10,0){8}{\circle*{2}}
\multiput(0,80)(10,0){8}{\circle*{2}} \multiput(0,90)(10,0){7}{\circle*{2}}
\multiput(0,100)(10,0){5}{\circle*{2}} \put(50,100){\circle*{3}}

\put(-8.5,5){$\lambda_2$}
\put(5.5,1.5){$\lambda_1$}
\put(-7,-5){\small$O$}
\put(-7,40){\small$F$}
\put(-7,50){\small$A$}
\put(-7,100){\small$B$}
\put(54,98){\small$C(10,5)$}
\put(22,16){\small$G$}
\put(51,45){\small$E$}
\put(78,74){\small$D$}

\small
\multiput(1,99)(10,0){6}{$\bf^1$}
\multiput(1,89)(10,0){5}{$\bf^3$} \put(51,89){$\bf^2$} \put(61,89){$\bf^1$}
\multiput(1,79)(10,0){4}{$\bf^6$} \put(41,79){$\bf^5$} \put(51,79){$\bf^4$} 
\put(61,79){$\bf^2$} \put(70,79){$\bf^1$}
\multiput(1,69)(10,0){3}{$\bf^{10}$} \put(31,69){$\bf^9$} \put(41,69){$\bf^8$} 
\put(51,69){$\bf^6$} \put(61,69){$\bf^4$} \put(68,69){$\bf^2$}
\multiput(1,59)(10,0){2}{$\bf^{15}$} \put(21,59){$\bf^{14}$} \put(31,59){$\bf^{13}$} 
\put(41,59){$\bf^{11}$} \put(51,59){$\bf^9$} \put(58,59){$\bf^6$}
\put(1,49){$\bf^{21}$} \put(11,49){$\bf^{20}$} \put(21,49){$\bf^{19}$} 
\put(31,49){$\bf^{17}$} \put(40.5,49){$\bf^{15}$} \put(47,49){$\bf^{12}$}
\put(1,39){$\bf^{25}$} \put(11,39){$\bf^{25}$} \put(21,39){$\bf^{23}$} 
\put(30.5,39){$\bf^{21}$} \put(36.5,39){$\bf^{18}$}  
\put(1,29){$\bf^{29}$} \put(11,29){$\bf^{28}$} \put(20,29){$\bf^{27}$} 
\put(26.5,29){$\bf^{24}$}  
\put(1,19){$\bf^{31}$} \put(11,19){$\bf^{31}$}  \put(17,20.4){$\bf^{29}$}  
\put(1,9){$\bf^{33}$}   \put(7,10){$\bf^{32}$} 
\put(0.5,2.5){$\bf^{33}$}
\end{picture}
\qquad\qquad\qquad
\begin{picture}(100,140)(0,0)
\put(0,0){\thicklines\line(1,0){100}}
\put(0,0){\thicklines\line(0,1){150}}
\put(0,0){\line(1,1){97}}
\put(0,0){\thicklines\vector(0,1){10}}
\put(0,0){\thicklines\vector(1,1){5}}
\put(50,140){\line(1,-1){47}}
\put(0,140){\line(1,0){50}}
\put(0,0){\circle{3}} \put(0,0){\circle*{2}}
\multiput(0,10)(10,0){2}{\circle{3}} \multiput(0,10)(10,0){2}{\circle*{2}}
\multiput(0,20)(10,0){3}{\circle{3}} \multiput(0,20)(10,0){3}{\circle*{2}} 
\multiput(0,30)(10,0){4}{\circle{3}} \multiput(0,30)(10,0){4}{\circle*{2}} 
\multiput(0,40)(10,0){5}{\circle{3}} \multiput(0,40)(10,0){5}{\circle*{2}}
\multiput(0,50)(10,0){4}{\circle{2}} \multiput(40,50)(10,0){2}{\circle*{2}}
\multiput(0,60)(10,0){3}{\circle{2}} \multiput(30,60)(10,0){4}{\circle*{2}}
\multiput(0,70)(10,0){2}{\circle{2}} \multiput(20,70)(10,0){6}{\circle*{2}}
\put(0,80){\circle{2}}  \multiput(10,80)(10,0){8}{\circle*{2}}
\multiput(0,90)(10,0){10}{\circle*{2}}
\multiput(0,100)(10,0){10}{\circle*{2}}
\multiput(0,110)(10,0){9}{\circle*{2}}
\multiput(0,120)(10,0){8}{\circle*{2}}
\multiput(0,130)(10,0){7}{\circle*{2}}
\multiput(0,140)(10,0){5}{\circle*{2}} \put(50,140){\circle*{3}}

\put(-8,5){$\lambda_2$}
\put(5.5,1.5){$\lambda_1$}
\put(-7,-5){\small$O$}
\put(-7,50){\small$A$}
\put(-7,80){\small$F$}
\put(-7,140){\small$B$}
\put(54,138){\small$C(10,9)$}
\put(40,34){\small$H$}
\put(50,45){\small$E$}
\put(92,87){\small$D$}
\put(29.5,45){\small$G$}

\small
\multiput(1,139)(10,0){6}{$\bf^1$}
\multiput(1,129)(10,0){5}{$\bf^3$} \put(51,129){$\bf^2$} \put(61,129){$\bf^1$}

\multiput(1,119)(10,0){4}{$\bf^6$} \put(41,119){$\bf^5$} \put(51,119){$\bf^4$} 
\put(61,119){$\bf^2$} \put(71,119){$\bf^1$}

\multiput(1,109)(10,0){3}{$\bf^{10}$} \put(31,109){$\bf^9$} \put(41,109){$\bf^8$} 
\put(51,109){$\bf^6$} \put(61,109){$\bf^4$} \put(71,109){$\bf^2$} \put(81,109){$\bf^1$}

\multiput(1,99)(10,0){2}{$\bf^{15}$} \put(21,99){$\bf^{14}$} \put(31,99){$\bf^{13}$} 
\put(41,99){$\bf^{11}$} \put(51,99){$\bf^9$} \put(61,99){$\bf^6$} \put(71,99){$\bf^4$} \put(81,99){$\bf^2$} \put(91,99){$\bf^1$}

\put(1,89){$\bf^{21}$} \put(11,89){$\bf^{20}$} \put(21,89){$\bf^{19}$} 
\put(31,89){$\bf^{17}$} \put(41,89){$\bf^{15}$} \put(51,89){$\bf^{12}$} \put(61,89){$\bf^{9}$}  \put(71,89){$\bf^{6}$} \put(81,89){$\bf^{4}$}  \put(88,89){$\bf^{2}$} 

\multiput(1,79)(10,0){2}{$\bf^{26}$} \put(21,79){$\bf^{24}$} \put(31,79){$\bf^{22}$} 
\put(41,79){$\bf^{19}$} \put(51,79){$\bf^{16}$}  \put(61,79){$\bf^{12}$} 
\put(71,79){$\bf^{9}$} \put(78,79){$\bf^{6}$} 

\put(1,69){$\bf^{32}$} \put(11,69){$\bf^{31}$} \put(21,69){$\bf^{30}$} 
\put(31,69){$\bf^{27}$}   \put(41,69){$\bf^{24}$}  \put(51,69){$\bf^{20}$}  \put(60,69){$\bf^{16}$}  \put(66.5,69){$\bf^{12}$} 
 
\put(1,59){$\bf^{37}$} \put(11,59){$\bf^{37}$} \put(21,59){$\bf^{35}$} \put(31,59){$\bf^{33}$} \put(41,59){$\bf^{29}$} \put(50,59){$\bf^{25}$} \put(57,59){$\bf^{20}$} 

\put(1,49){$\bf^{43}$}  \put(11,49){$\bf^{42}$}   \put(21,49){$\bf^{41}$}   \put(31,49){$\bf^{38}$}   \put(40.5,49){$\bf^{35}$}   \put(47,49){$\bf^{30}$}    

\put(1,39){$\bf^{47}$}   \put(11,39){$\bf^{47}$}  \put(21,39){$\bf^{45}$}  \put(30,39){$\bf^{43}$}  \put(37,39){$\bf^{39}$}  

\put(1,29){$\bf^{51}$}  \put(11,29){$\bf^{50}$}  \put(20,29){$\bf^{49}$}  \put(27,29){$\bf^{46}$}  

\put(1,19){$\bf^{53}$}  \put(10.5,19){$\bf^{53}$}  \put(17,19.5){$\bf^{51}$}  

\put(1,9){$\bf^{55}$}  \put(8.,9.5){$\bf^{54}$}  

\put(0.5,2.5){$\bf^{55}$}
\end{picture}
\end{center}
\medskip

{\footnotesize Figure 2. The weights of the representation $R_{10\lambda_1+5\lambda_2}$ in the domains $D1$, $D0$ and $D(-1)$ are marked, 
respectively, with black dots,   circles and   encircled black dots. The region $P_X\cup P_Y$ 
has perimeter $ABCDEA$, while $P_Z$ is contained in $FGOF$. The number over a weight means 
its multiplicity.}
\vskip2mm

{\footnotesize Figure 3.  Weights of the representation $R_{10\lambda_1+9\lambda_2}$. The region $P_X\cup P_Y$ is bounded by $AFBCDEGA$, $P_Z$ is 
into $FGHOAF$ and the triangle $T$ is $FGAF$.}
\end{figure}

\end{itemize}

\begin{figure}
\begin{center}
\setlength{\unitlength}{.7mm}

\begin{picture}(80,140)(0,0)
\put(0,0){\thicklines\line(1,0){80}}
\put(0,0){\thicklines\line(0,1){140}}
\put(0,0){\line(1,1){80}}
\put(0,0){\thicklines\vector(0,1){10}}
\put(0,0){\thicklines\vector(1,1){5}}
\put(25,125){\line(1,-1){55}}
\put(0,125){\line(1,0){25}}

\put(5,5){\circle{3}} \put(5,5){\circle*{2}}
\multiput(5,15)(10,0){2}{\circle{3}} \multiput(5,15)(10,0){2}{\circle*{2}}
\multiput(5,25)(10,0){3}{\circle{2}} 
\multiput(5,35)(10,0){4}{\circle{2}} 
\multiput(5,45)(10,0){5}{\circle{2}} 
\multiput(5,55)(10,0){4}{\circle{2}} \multiput(45,55)(10,0){2}{\circle*{2}}
\multiput(5,65)(10,0){3}{\circle{2}} \multiput(35,65)(10,0){4}{\circle*{2}}
\multiput(5,75)(10,0){2}{\circle{2}} \multiput(25,75)(10,0){6}{\circle*{2}}
\put(5,85){\circle{2}}  \multiput(15,85)(10,0){6}{\circle*{2}}
\multiput(5,95)(10,0){6}{\circle*{2}}
\multiput(5,105)(10,0){5}{\circle*{2}}
\multiput(5,115)(10,0){4}{\circle*{2}}
\multiput(5,125)(10,0){2}{\circle*{2}} \put(25,125){\circle*{3}}

\put(-8,5){$\l_2$}
\put(5,.7){\scriptsize${\l_1}$}
\put(-6,-5){\small$O$}
\put(-6,23){\small$A$}
\put(-6,90){\small$F$}
\put(-6,125){\small$B$}
\put(27,123){\small$C(5,10)$}
\put(26,21){\small$E$}
\put(78,73){\small$D$}
\put(46,41){\small$G$}

\small
\multiput(5,125)(10,0){3}{$\bf^1$} 

\put(5,115){$\bf^3$} \put(15,115){$\bf^3$} \put(25,115){$\bf^2$} \put(35,115){$\bf^1$} 
 
 \put(5,105){$\bf^6$}   \put(15,105){$\bf^5$}  \put(25,105){$\bf^4$}  \put(35,105){$\bf^2$}  \put(45,105){$\bf^1$} 
 
  \put(5,95){$\bf^9$}   \put(15,95){$\bf^8$}  \put(25,95){$\bf^6$}  \put(35,95){$\bf^4$}  \put(45,95){$\bf^2$}  \put(55,95){$\bf^1$}
  
\put(5,85){$\bf^{12}$} \put(15,85){$\bf^{11}$} \put(25,85){$\bf^9$} \put(35,85){$\bf^6$} \put(45,85){$\bf^4$} \put(55,85){$\bf^2$} \put(65,85){$\bf^1$} 

\put(5,75){$\bf^{15}$} \put(15,75){$\bf^{14}$} \put(25,75){$\bf^{12}$} \put(35,75){$\bf^{9}$} \put(45,75){$\bf^{6}$} \put(55,75){$\bf^{4}$} \put(65,75){$\bf^{ 2}$} \put(74,75){$\bf^{1}$} 

\put(5,65){$\bf^{18}$} \put(15,65){$\bf^{17}$} \put(25,65){$\bf^{15}$} \put(35,65){$\bf^{12}$} \put(45,65){$\bf^{9}$} \put(55,65){$\bf^{6}$} \put(64,65){$\bf^{4}$} 

\put(5,55){$\bf^{21}$} \put(15,55){$\bf^{20}$} \put(25,55){$\bf^{18}$} \put(35,55){$\bf^{15}$} \put(45,55){$\bf^{12}$} \put(55,55){$\bf^{9}$} 

\put(5,45){$\bf^{24}$} \put(15,45){$\bf^{23}$} \put(25,45){$\bf^{21}$} \put(35,45){$\bf^{18}$} \put(43,45){$\bf^{15}$} 

\put(5,35){$\bf^{27}$} \put(15,35){$\bf^{26}$} \put(25,35){$\bf^{24}$} \put(33,35){$\bf^{21}$} 

\put(5,25){$\bf^{30}$}  \put(15,25){$\bf^{29}$}  \put(23,25){$\bf^{27}$} 

\put(5,15){$\bf^{32}$} \put(13,15){$\bf^{31}$} 

\put(3,5){$\bf^{33}$}

\end{picture}
\qquad\qquad\qquad
\begin{picture}(100,140)(0,0)
\put(0,0){\line(1,0){100}}
\put(0,0){\line(0,1){150}}
\put(0,0){\line(1,1){97}}
\put(0,0){\thicklines\vector(0,1){10}}
\put(0,0){\thicklines\vector(1,1){5}}
\put(50,140){\line(1,-1){47}}
\put(0,140){\line(1,0){50}}

\put(0,0){\circle*{2}}
\multiput(0,10)(10,0){2}{\circle*{2}}
\multiput(0,20)(10,0){3}{\circle*{2}} 
\multiput(0,30)(10,0){4}{\circle*{2}} 
\multiput(0,40)(10,0){5}{\circle*{2}}
\multiput(0,50)(10,0){6}{\circle*{2}}
\multiput(0,60)(10,0){7}{\circle*{2}}
\multiput(0,70)(10,0){8}{\circle*{2}} 
\multiput(0,80)(10,0){9}{\circle*{2}}
\multiput(0,90)(10,0){10}{\circle*{2}}
\multiput(0,100)(10,0){10}{\circle*{2}}
\multiput(0,110)(10,0){9}{\circle*{2}}
\multiput(0,120)(10,0){8}{\circle*{2}}
\multiput(0,130)(10,0){7}{\circle*{2}}
\multiput(0,140)(10,0){6}{\circle*{2}}

\put(0,70){\line(1,-1){20}}
\put(20,50){\line(1,0){30}}

\multiput(18.5,48.5)(10,0){4}{\framebox(3,3){ }}
\multiput(8.5,58.5)(10,0){5}{\framebox(3,3){ }}
\multiput(-1.5,68.5)(10,0){6}{\framebox(3,3){ }}
\multiput(8.5,78.5)(10,0){5}{\framebox(3,3){ }}
\multiput(-1.5,88.5)(10,0){6}{\framebox(3,3){ }}
\multiput(8.5,98.5)(10,0){5}{\framebox(3,3){ }}
\multiput(18.5,108.5)(10,0){4}{\framebox(3,3){ }}
\multiput(28.5,118.5)(10,0){3}{\framebox(3,3){ }}
\multiput(38.5,128.5)(10,0){2}{\framebox(3,3){ }}
\multiput(48.5,138.5)(10,0){1}{\framebox(3,3){ }}

\put(-8,5){$\l_2$}
\put(5,1.5){$\l_1$}
\put(-6,-5){\small$O$}
\put(-7,80){\small$X$}
\put(13,44){\small$(m,n)$}
\put(50,144){$\small(p,q)$}

\end{picture}
\end{center}
\medskip

{\footnotesize Figure 4.  Weights of the representation $P_{5\l_1+10\l_2}$. The region $P_X\cup P_Y$ is bounded by $AFBCDGEAF$, $P_Z$ is into 
$FGEOAF$ and the cuadrilateral $K$ is $FGEAF$.}
\medskip

{\footnotesize Figure 5. The multiplicity of $m\l_1+n\l_2$ in the representation 
$R_{p\l_1+q\l_2}$ is given by the number of weights marked with a square.  By parity, the weight signaled with $X$ is excluded.}
\end{figure}

\section{An application}
The recurrence relations of the previous sections provide useful information on the multiplicities and, in fact, 
can be used to devise some simple rules to compute the multiplicity of any desired weight on a given representation.  As an example, let us take the case of $m$ even and consider  a situation with $(m,n)$ and $(p,q)$ as given in the figure 5. We can write the first recurrence relation (\ref{recrel}) as
\bdm
\mu_{p,q}(m,n)-\mu_{p,q-1}(m,n)=y_{p,q}(m,n)+\mu_{p-2,q}(m,n)-\mu_{p-2,q-1}(m,n)
\edm
and iterating we find
\beqrn
\mu_{p,q}(m,n)&=&y_{p,q}(m,n)+y_{p-2,q}(m,n)+\cdots+y_{2,q}(m,n)+y_{0,q}(m,n)\\
&+&y_{p,q-1}(m,n)+y_{p-2,q-1}(m,n)+\cdots+y_{2,q-1}(m,n)+y_{0,q-1}(m,n)\\
&+&y_{p,q-2}(m,n)+y_{p-2,q-2}(m,n)+\cdots+y_{2,q-2}(m,n)+y_{0,q-2}(m,n)+\cdots
\eeqrn
so that finally 
\bdm
\mu_{p,q}(m,n)=\sum_{\beta\in R} y_\beta (m,n)
\edm 
where the sum is over to the set $R$ of weights marked in the figure. Thus, $\mu_{p,q}(m,n)$ 
can be obtained by simply counting the number of points in $R$ except those in the vertical 
axis with opposite parity of $q$ and $n$, which have $y_\beta (m,n)=0$. This rule can be used 
to obtain some explicit formulae. For instance, for the case $p$ and $q$ even and 
$q\leq {p}/{2}$, one finds that the multiplicities of the weights on borders of the diagram 
are given by
\beqrn
&&\mu_{p,q}(p+q-2s,0)=(s+1)^2{\hspace{3.5cm}\rm for\ \ \ }0\leq s\leq \frac{q}{2}\\
&&\mu_{p,q}(p-2s,0)=(\frac{q}{2}+1)^2+s(q+1){\hspace{2.2cm}\rm for\ \ \ }0\leq s\leq \frac{p-q}{2}\\
&&\mu_{p,q}(2s,0)=\mu_{p,q}(0,0)-s^2{\hspace{3.9cm}\rm for\ \ \ }0\leq s\leq \frac{q}{2}
\eeqrn
and
\beqrn
&&\mu_{p,q}(0,\frac{p}{2}+q-s)=\frac{(s+1)(s+2)}{2}{\hspace{2.94cm}\rm for\ \ \ }0\leq s\leq q\\
&&\mu_{p,q}(0,\frac{p}{2}-s)=\frac{(s+1)(s+2)}{2}+s(q+1){\hspace{1.7cm}\rm for\ \ \ }0\leq s\leq \frac{p}{2}-q\\
&&\mu_{p,q}(0,s)-\mu_{p,q}(0,s+1)=s+1-\theta(s+1){\hspace{1.35cm}\rm for\ \ \ }0\leq s\leq q-1
\eeqrn
where $\theta(r)$ is one (zero) for $r$ even (odd). The combination of these formulas with the recurrence relation (\ref{eq:rec2}) 
can  be used as another alternative to compute the multiplicities of the inner weights.

\section*{Acknowledgement}
J.F.N. acknowledges financial support from MTM2012-33575 project, SGPI-DGICT(MEC), Spain.

\section*{Appendix}
We list here some generating functions of multiplicities of low-lying weights of $C_2$ up to $m+n=4$ as obtained by computing the corresponding integrals in formula (\ref{amni}).
\beqrn
A_{1,0}(t_1,t_2)&=&\frac{t_1}{(1 - t_1^2)^2 (1 - t_2)^2}\\
[2pt]
A_{0,1}(t_1,t_2)&=&\frac{t_1^2 + t_2}{(1 - t_1^2)^2 (1 - t_2)(1 - t_2^2)}\\
[2pt]
A_{2,0}(t_1,t_2)&=&\frac{t_2^2+t_1^2(1+t_2-t_2^2)}{(1-t_1^2)^2(1-t_2)^2(1+t_2)}\\
[2pt]
A_{1,1}(t_1,t_2)&=&\frac{t_1^3(1-t_2)+t_1t_2}{(1-t_1^2)^2(1-t_2)^2}\\
[2pt]
A_{0,2}(t_1,t_2)&=&\frac{t_1^2t_2+t_2^2+t_1^4(1-t_2^2)}{(1-t_1^2)^2(1-t_2)^2(1+t_2)}\\
[2pt]
A_{3,0}(t_1,t_2)&=&\frac{t_1t_2^2+t_1^3(1-t_2^2)}{(1-t_1^2)^2(1-t_2)^2}\\
[2pt]
A_{2,1}(t_1,t_2)&=&\frac{t_2^3+t_1^4(1-t_2^2)+t_1^2t_2(1+t_2-t_2^2)}{(1-t_1^2)^2(1-t_2)^2(1+t_2)}\\
[2pt]
A_{1,2}(t_1,t_2)&=&\frac{t_1^5(1-t_2)+t_1^3 t_2(1-t_2)+t_1t_2^2}{(1-t_1^2)^2(1-t_2)^2}\\
[2pt]
A_{0,3}(t_1,t_2)&=&\frac{t_1^2t_2^2+t_2^3+t_1^6(1-t_2^2)+t_1^4t_2(1-t_2^2)}{(1-t_1^2)^2(1-t_2)^2(1+t_2)}\\
[2pt]
A_{4,0}(t_1,t_2)&=&\frac{t_2^4 + t_1^4(1 - t_2)(1 + t_2)^2 + 
    t_1^2t_2^2(1 + t_ 2- t_2^2)}{(1 - t_1^2)^2(1 - t_2)^2(1 + t_2)}\\
[2pt]
A_{3,1}(t_1,t_2)&=&\frac{t_1^5(1 - t_2) + t_1t_2^3 + t_1^3t_2(1 - t_2^2)}{(1-t_1^2)^2(1 - t_2)^2}\\
[2pt]
A_{2,2}(t_1,t_2)&=&\frac{t_2^4+t_1^6(1-t_2^2)+t_1^4t_2(1-t_2^2)+t_1^2t_2^2(1+t_2-t_2^2)}               {(1-t_1^2)^2(1-t_2)^2(1+t_2)} \\
[2pt]
A_{1,3}(t_1,t_2)&=&\frac{t_1^7(1 - t_2) + t_1^5t_2(1 - t_2) +t_1^3t_2^2(1 - t_2) + 
    t_1t_2^3}{(1 - t_1^2)^2(1 - t_2)^2}\\
[2pt]
A_{0,4}(t_1,t_2)&=&\frac{t_1^2t_2^3 + t_2^4 +t_1^8(1 - t_2^2) + t_1^6t_2(1 - t_2^2) + 
   t_1^4t_2^2(1 - t_2^2)}{(1 - t_1^2)^2(1 - t_2)^2(1 + t_2)} \\
[2pt]
\eeqrn


\begin{thebibliography}{AA99}

\bibitem{wi37}  Wigner E P 1937  {\it Phys. Rev.} {\bf 51} 106--119
\bibitem{fr54}  Freudenthal H 1954 {\it Proc. Nederl. Akad. Wet.} {\bf A57}  369--376, 487--491
\bibitem{ra62} Racah G 1964 {\it Lectures on Lie groups}, in  {\it Group theoretical concepts
and methods in elementary particle physics} (Lectures Istanbul Summer
School Theoret. Phys., 1962) (New York: Gordon and Breach) pp 1-36
\bibitem{ko59} Kostant B 1959 {\it Trans. Amer. Math. Soc.} {\bf 93}  53-73  
\bibitem{mopa82}  Moody R V and   Patera J 1982 {\it Bull. Amer. Math. Soc.} {\bf 7}  237-242 
\bibitem{nfp14} Fern\'andez--N\'{u}\~{n}ez J,  Garc\'{\i}a--Fuertes W and   Perelomov A M 2014 {\it J. Phys. A: Math. Theor.} {\bf 47} 145202 
\bibitem{ca71} Calogero F 1971 {\it J. Math. Phys.} {\bf 12}  419--436 
\bibitem{su72}  Sutherland B 1972 {\it Phys. Rev.}  {\bf A4} 2019--2021  
\bibitem{mo75}   Moser J 1975 {\it Adv. Math.} {\bf 16} 197--220  
\bibitem{op76}  Olshanetsky  M A and Perelomov A M {\it Invent. Math.} {\bf 37}  93--108  
\bibitem{ps79}   Patera J and   Sharp R T 1979 {\it Lecture Notes in Physics} {\bf 94} 175-183 (Berlin: Springer)
\bibitem{ow07}  Okeke N and   Walton M A 2007 {\it J. Phys. A: Math. Theor.} {\bf 40} 8873-8901 
\bibitem{bgw68}   Biedenharn L C,   Gruber B and   Weber H J 1968 {\it Proc.   R. Ir. Acad.} {\bf  A 67} 1--14
\bibitem{ov90}  Onishchik A L and    Vinberg E B 1990  {\it Lie Groups and Algebraic Groups} (Berlin: Springer)
\bibitem{dok}  Dokovi\'c D Z 1995 {\it Indag. Mathem. N. S.} {\bf 6} 145--151

\end{thebibliography}
\end{document}